\documentclass[twocolumn,prb,showpacs]{revtex4}
\usepackage{curvesls,graphicx,amssymb,longtable,exscale,amsmath,epsfig}
\usepackage[all]{xy}

\begin{document}
\draft                  %do not suppress PACS number
\preprint{Phys. Rev. B: April 20 version}
\title{Role of the exchange and correlation potential into calculating the x-ray absorption spectra of
half-metallic alloys: the case of
Mn and Cu K-edge XANES in Cu$_2$MnM (M = Al, Sn, In) Heusler
alloys.}
\author{K. Hatada$^{1,2,*}$ and J. Chaboy$^1$}
\affiliation{$^1$Instituto de Ciencia de Materiales de Arag\'on,
CSIC-Universidad de Zaragoza, 50009 Zaragoza, Spain, \\
$^2$Laboratori Nazionali di Frascati, LNF-INFN, 00044 Frascati, Italy.}
\date{\today}

\begin{abstract}

This work reports a theoretical study of the x-ray absorption
near-edge structure spectra at both the Cu and the Mn K-edge in
several Cu$_2$MnM (M= Al, Sn and In) Heusler alloys. Our results
show that {\it ab-initio} single-channel multiple-scattering
calculations are able of reproducing the experimental spectra.
Moreover, an extensive discussion is presented concerning the role
of the final state potential needed to reproduce the experimental
data of these half-metallic alloys. In particular, the effects of
the cluster-size and of the exchange and correlation potential
needed in reproducing all the experimental XANES features are
discussed.

\end{abstract}

\maketitle

\section{INTRODUCTION}

X-ray absorption spectroscopy (XAS) constitutes nowadays an
outstanding structural tool. The extended x-ray absorption fine
structure (EXAFS) part of the spectrum is commonly used to determine
the local environment around a selected atomic species in a great
variety of systems \cite{general}. Moreover, the near-edge part of
the absorption spectrum (XANES) becomes an incomparable
stereochemical probe because its high sensitivity to the bonding
geometry. This capability is due to the low kinetic energy of the
photoelectron that favors the contribution of multiple scattering
processes. By this reason, a great effort has been devoted in the
last decade to obtain structural determinations from XANES,
including bond-angles information. However, the interpretation and
the {\it ab-initio} computation of XANES spectra are not so
straightforward as for EXAFS.

XANES computation requires sophisticated simulation tools
\cite{g2,g1,CONTINUUM,FEFF8,GNXAS,Joly2,MXAN}. Most of these {\it
ab-initio} codes are built within the one-electron
multiple-scattering (MS) framework \cite{Lee} and by using the so
called muffin-tin approximation. More recently, several works appear
reporting XANES computations performed within a non-muffin-tin
approach \cite{Joly2,Joly1,Hatada2} and the promising implementation of the
multi-channel MS theory \cite{Natoli2,Kruger}. 
Despite the above progresses, the construction of the scattering
potential still remains an open problem. Typically,
the standard {\it ab initio} XANES calculations are performed in the
framework of the muffin-tin approximation with the addition of an
exchange-correlation potential (ECP). In the case of complex ECP
potentials, the imaginary part also accounts for the damping of the
excited photoelectron. At present, the energy dependent
Hedin-Lundqvist (HL) complex potential \cite{Hedin,Hatada} is the
most widely used into {\it ab initio} XANES calculations
\cite{Rehr}. However, it has been reported that in several cases the
energy dependent Dirac-Hara (DH) exchange potential shows the best
agreement to the experimental data \cite{Gunnella,Itrio,Ti,Briois},
although this assignment is not free of controversy
\cite{SiO2,Rose,Cabaret,Cabaret2}. As a result, the choice of the
ECP among the usual scattering potentials set, ranging from real
X$_\alpha$ to complex Hedin-Lundqvist, becomes one the most
important steps into obtaining a good reproduction of the
experimental spectra. The experience accumulated through the
calculation of the absorption cross section for different systems
suggests that the HL-ECP offers a good performance in the case of
metals, while for ionic and covalent systems the use of DH seems to
be more adequate\cite{Gunnella,Itrio,Ti,Briois,Modrow}.
However, the variety of reported results indicates that this choice
is material specific and, unfortunately, prevents one of {\it
a-priori} fixing the ECP for each particular case. In this work we
have faced this problem in connection to the case of half-metallic
alloys. These systems present a peculiar behavior associated to the
metal and insulator or semiconductor character of the spin
subsystems.

The Cu$_2$MnM (M= Al, Sn and In) Heusler alloys are interesting
magnetic systems because they possess localized magnetic moments
although they are all metallic. The magnetic, electrical and
structural properties in these systems are known to be strongly
dependent on both the conduction electron concentration and the
chemical order \cite{Webster,Webster2,c1_Oxley,c2_Soltys,Ishida2}.
Despite the importance of getting a deeper insight in the electronic
structure and the chemical bonding mechanism in these compounds, few
amount of research has been performed regarding the magnetic
polarization of the conduction electrons and its role into the
magnetic ordering. Recently, Uemura {\it et al.} have reported the
results of a combined XAS and x-ray magnetic circular dichroism
(XMCD) study performed at the K-absorption edge of both Mn and Cu in
these Cu$_2$MnM Heusler alloys \cite{Uemura}. Theoretical
 calculations have been made on the basis of
photoelectron full multiple scattering (FMS) theory \cite{Fujikawa},
and by using the Madelung potential taken from the band calculation
for Cu$_2$MnAl performed by Ishida {\it et al.} \cite{Ishida2}.
According to Uemura {\it et al.} the spin-orbit interaction and the
exchange scattering are the main factors for the calculation. 
Thus, even when they used a non-local exchange potential
that can directly take the spin-polarization effects into account,
correlation-potential was not included. In addition, only 51 atoms,
5 {\AA}, were included to built up the cluster used for the
computations. The fact that computations were
performed for such a small cluster and without correlation potential
poses serious concerns regarding the reliability of the reached
conclusions.

In this work, we present detailed {\it ab initio} computations of
the Mn and Cu K-edge XANES spectra in the Cu$_2$MnM (M= Al, Sn and
In) Heusler alloys. These calculations of the absorption spectra
have been made within the multiple-scattering framework. Special
attention has been paid to establish the improvements obtained by
using different treatments of the exchange-correlation part in the
final state potential, intending to obtain the best agreement both
in energy and intensity, between the experimental and theoretical
XANES spectra.

\section{COMPUTATIONAL METHODS}

The computation of the XANES spectra was carried out using the
multiple-scattering code CONTINUUM \cite{CONTINUUM,MXAN} based on
the one-electron full-multiple-scattering theory \cite{Natoli2,Lee}.
A complete discussion of the procedure can be found in
Ref.\cite{Hatada} and Ref.\cite{Chaboy1}. Computations were made in
parallel mode by using the MPI library \cite{MPI}.

The potential for the different atomic clusters was approximated by
a set of spherically averaged muffin-tin (MT) potentials built by
following the standard Mattheis' prescription \cite{Mattheis}. The
muffin-tin radii were determined following the Norman's criterion
and by imposing an overlapping factor ranging from 1$\%$ to 10$\%$
\cite{Norman}. We have verified that the factor chosen within this
range does not affect the spectral shape and only slight differences
are found in the first 5 eV of the computed spectra. After
convolution of the spectra to account for both the core-hole
lifetime and the experimental resolution these differences are not
significant.

The Coulomb part of each atomic potential was generated using charge
densities from atomic code of non-local selfconsistent Dirac-Fock
code \cite{Desclaux,FEFF8}. The atomic orbitals were chosen to be
neutral for the ground state potential. Two different approximations
were tested for the excited state potential: i) the same potential
as for the ground state was used; ii) relaxed Z+1 approximation
\cite{Lee2}. During the present calculations we have found that the
screened and relaxed Z+1 option leads to the best performance into
simulating the experimental absorption spectra at both the Mn and Cu
K-edge. Finally, we have tested three different choices for the
exchange and correlation part of the final state potential:
X$_\alpha$, the energy dependent Hedin-Lundqvist (HL) complex
potential and the energy-dependent Dirac-Hara (DH) exchange
potential.

The computed spectra have been compared to the experimental XANES
data reported Uemura {\it et al.} \cite{Uemura}. In all the cases,
the calculated theoretical spectra have been further convoluted with
a Lorentzian shape function $\Gamma$= 1.5 eV to account for both the
core-hole lifetime \cite{Krause} and the experimental resolution.

\section{RESULTS AND DISCUSSION}

\begin{center}
\begin{figure}
\includegraphics[scale=1.0]{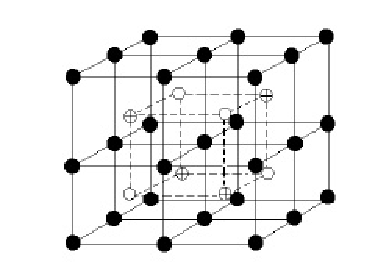}
\caption{The stoichiometric Cu$_2$MnM unit cell: Cu (black circles),
Mn (crossed circles) and M (open circles). \label{f1}}
\end{figure}
\end{center}

{\it Ab-initio} XANES calculations have been performed at both the
Mn and Cu K-edge in the case of Cu$_2$MnAl, Cu$_2$MnIn and
Cu$_2$MnSn. These Heusler alloys show the L2$_1$-type (space group
{\it Fm3m}) ordered structure illustrated in Fig.~\ref{f1}. The unit
cell contains eight of the small body-centred cubic cells that can
be regarded as four interpenetrating fcc sublattices.  There are
four kinds of atomic positions: Cu atoms occupy (1/4, 1/4, 1/4) and
(3/4, 3/4, 3/4) positions, Mn atoms are at (1/2, 1/2, 1/2), while
the M (Al, In, Sn) atoms occupy (0, 0, 0) position. The lattice
parameter is a = 5.957 {\AA} for Cu$_2$MnAl \cite{Soltys}; a = 6.166
{\AA} for Cu$_2$MnSn \cite{Uhl}; and a = 6.217 {\AA} for Cu$_2$MnIn
\cite{Webster3}.

\begin{center}
\begin{figure}
\includegraphics[scale=0.35]{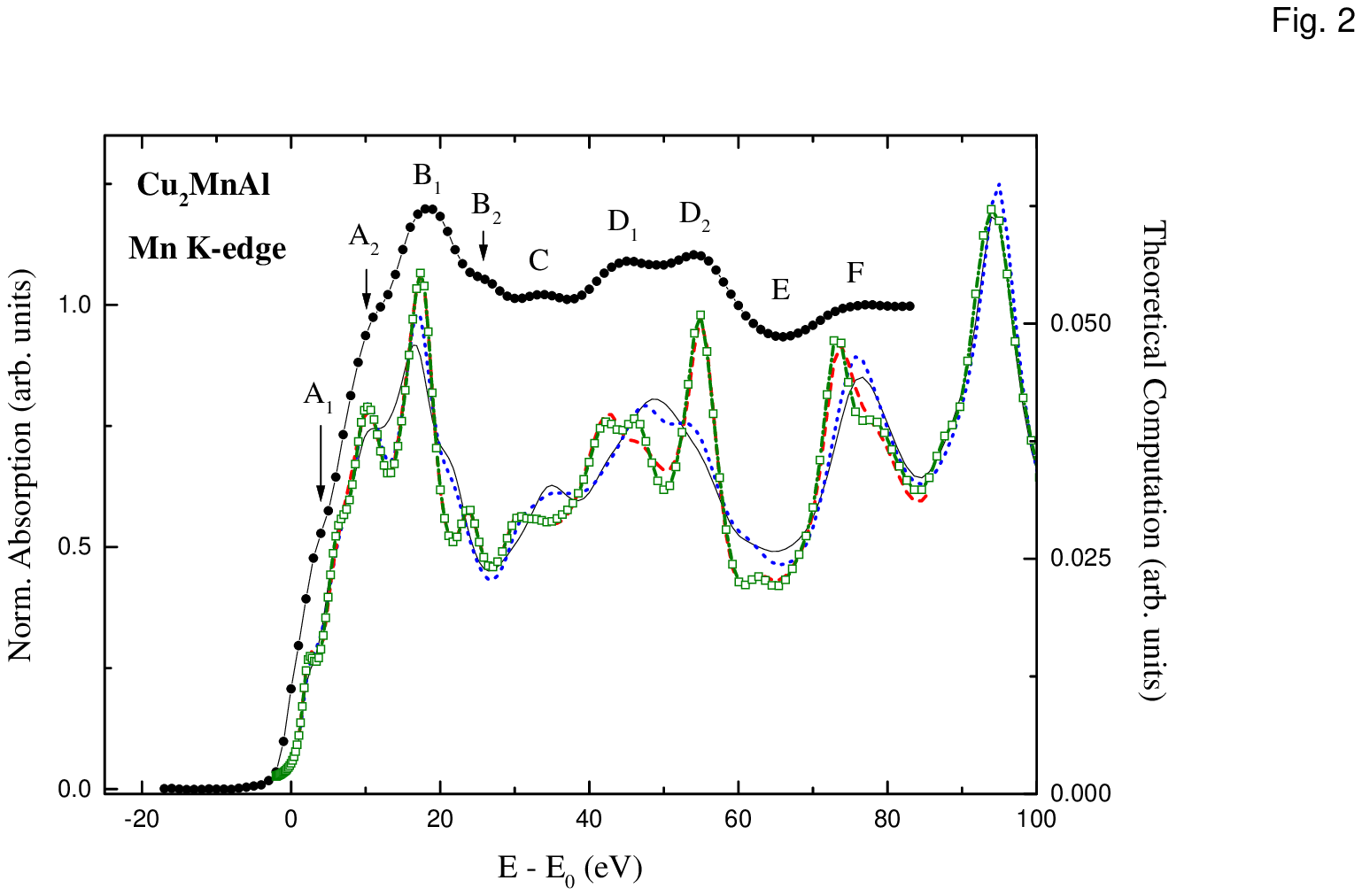}
\caption{(color online) Comparison between the experimental Mn
K-edge of Cu$_2$MnAl ($\bullet$) and the MS computations performed
for different cluster sizes covering up to 7.3 {\AA} around the
absorbing Mn: 59 (black, solid line), 65 (blue, dotted line), 113
(red, dashed line) and 137 atoms (green, $\circ$). \label{f2}}
\end{figure}
\end{center}

First step of the calculations was deserved to determine the size of
the cluster needed to reproduce all the spectral structures present
in the experimental XANES spectra. To this end we have performed the
computation of the Mn K-edge XANES in the case of Cu$_2$MnAl by
increasing progressively the number of atoms in the built-up
cluster. In this way we have computed the Mn K-edge absorption for
clusters containing: 59 atoms, i.e. including contributions from
neighbouring atoms located within the first 5.2 {\AA} around
photoabsorbing Mn; 65 (6 {\AA}), 113 (6.7 {\AA}) and 137 (7.3 {\AA})
atoms. The computations, shown in Fig.~\ref{f2}, have been performed
by using the real part of the HL ECP and with {\it l}$_{max}$ = 4.
The computed spectra show the dependence of both the energy position
and the relative intensity of the spectral features as the size of
the cluster increases. In the case of the smallest clusters, 59 and
65 atoms, several spectral features are not reproduced by the
calculation. In particular, the B$_2$ peak ( E $\sim$ 26 eV),
located at the high energy side of the main B$_1$ peak is not
reproduced. Moreover, for these clusters the computation yields a
broad resonance at $\sim$ 50 eV, whilst the experimental spectrum
exhibits two well defined peaks D$_1$ and D$_2$. By contrast, the
computations performed for bigger clusters, 113 and 137 atoms,
account for the previously missed structures. In addition, no
differences are found within the first 60 eV of the absorption
spectrum between computations performed for clusters including
contributions from atoms located within the first 6.7 {\AA} and 7.3
{\AA} around the photoabsorber. This result indicates that the
addition of further coordination shells does not contribute
significantly to the XANES spectrum. Accordingly, all
the calculations reported henceforth have been obtained by using 169
atoms to generate the scattering potential but only the scattering
contributions of the first 137 atoms ($\sim$ 6.7 {\AA}) are
computed.

Next, we have investigated the maximum angular momentum quantum
number, {\it l}$_{max}$, needed to account for the experimental
spectrum in the first 100 eV of the absorption spectrum. The choice
of {\it l}$_{max}$ = 3 and 4 does not affect the result of the
computations in the first 30 eV of the spectrum above the edge.
However, for higher energies both the energy position and intensity
of the computed features clearly differ. By increasing {\it
l}$_{max}$ from 4 to 5, the computed spectra do not change in the
simulated spectra in the energy interval of interest, up to $\simeq$
80 eV above the edge. Therefore, we have fixed {\it l}$_{max}$ = 4
for all the calculations.

\begin{center}
\begin{figure}
\includegraphics[scale=0.65]{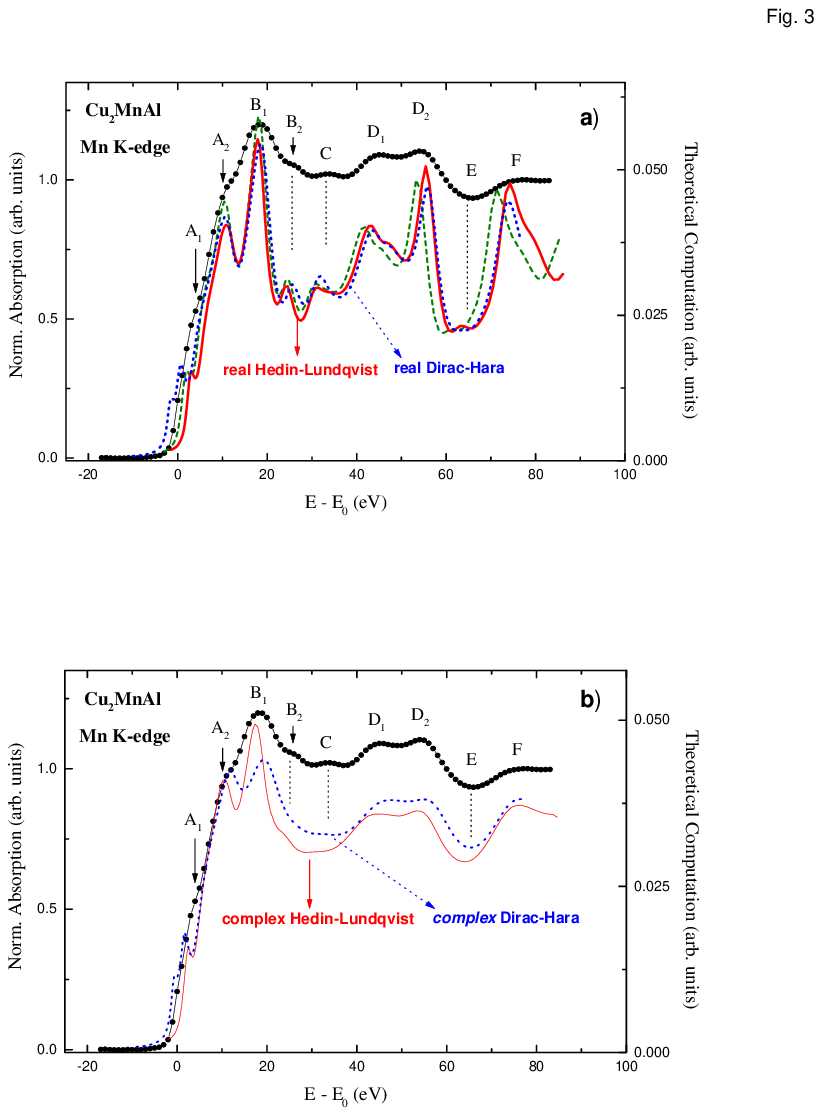}
\caption{(color online) Comparison between the experimental Mn
K-edge of Cu$_2$MnAl ($\circ$) and the MS computations performed for
different ECP potentials: a) X$\alpha$ (green dashed line), real HL
(red, solid line) and real DH (blue, dotted line); b) complex HL
(red, solid line) and {\it complex} DH (blue, dotted line) (see text
for details). \label{f4}}
\end{figure}
\end{center}

Based on the above results we have fixed both the cluster size and
{\it l}$_{max}$. Then, we have investigated the role of the
different exchange and correlation potentials (ECP) into correctly
reproduce the experimental spectra. We have computed the Mn K-edge
XANES of Cu$_2$MnAl by using the X$\alpha$ and the energy dependent
Hedin-Lundqvist (HL) and Dirac-Hara (DH) ECP potentials
\cite{Hatada}. The HL is a complex potential in which its imaginary
accounts for the inelastic losses of the photoelectron. Therefore,
we have also made computations by using only the real part of the HL
ECP (hereafter, real HL) and, in addition, we have built a {\it
"complex"} DH one by adding to it the imaginary part of the HL one.
The results of the computations are shown in Fig.~\ref{f4}. For the
sake of clarity, the comparison between the theoretical spectra and
the experimental one has been divided in two panels, one for the
real potentials (a) and the second for the complex ones (b). In the
case of the real potentials, the intensity of the spectral features
is not in agreement with the experimentally observed, as expected
because the photoelectron damping is not taken into account.
However, its inspection is rather useful as it allows one to
determine that the computations reproduce all the spectral features
shown by the experimental spectrum. To this respect, the agreement
between the computations and the experimental signal is noticeable
as the calculations reproduce the complex profile at the threshold
(A$_1$ and A$_2$ features), the main peaks (B$_1$, D$_1$, D$_2$ and
F) and the two tiny structures, B$_2$ and C. The performance of the
different potentials is similar although in the case of both
X$\alpha$ and the real HL the calculated absorption maxima falling
short of the observed ones. The addition of the imaginary part of
the HL potential improves the calculations. Good agreement between
the computations and the experimental spectrum is obtained for both
complex Hedin-Lundqvist ECP and by adding to the Dirac-Hara (DH)
exchange the imaginary part of the HL (hereafter complex
Dirac-Hara). As shown in Fig.~\ref{f4}b) all the experimental
features are properly accounted by the computations. Complex HL
computation yields a better reproduction of the intensity ratio
among the different spectral features, specially regarding the main
peak B$_1$ and both A$_1$ and B$_2$ features. By contrast complex DH
shows a better reproduction of the relative energy position among
the peaks. The experimental energy separation between the main
absorption peak B$_1$ and the negative E deep is $\Delta$E = 48 eV,
while computation yields 47.6 eV and 46.2 for complex DH and HL,
respectively.

\begin{center}
\begin{figure}
\includegraphics[scale=0.65]{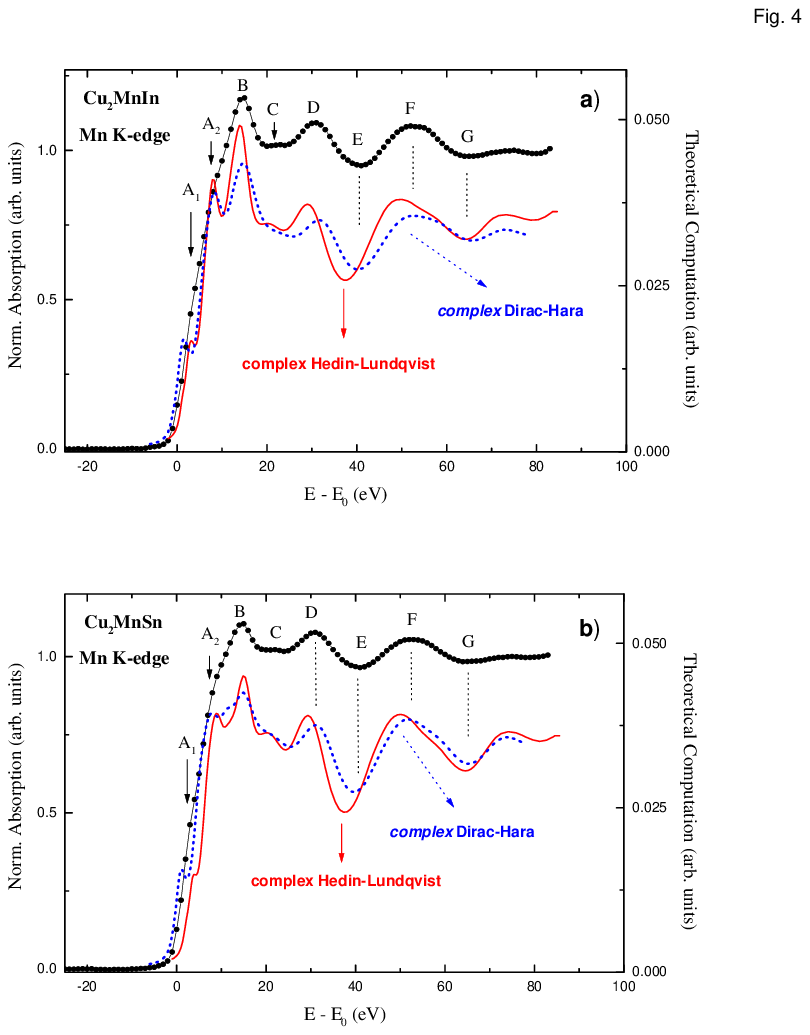}
\caption{(color online) Comparison between the MS computations
performed by using both complex HL (red, solid line) and {\it
complex} DH (blue, dotted line) and the experimental Mn K-edge
($\circ$) of: a) Cu$_2$MnIn  and b) Cu$_2$MnSn. \label{f5}}
\end{figure}
\end{center}

We have performed the same class of computations for both Cu$_2$MnIn
and Cu$_2$MnSn systems. In all cases we have used {\it l}$_{max}$ =
4 and the potential was created by using a 169 atoms cluster to
calculate the scattering contributions of the first 137 atoms. In
this way, we account for the contributions from atoms located within
the first 6.9 {\AA} and 6.95 {\AA} around Mn for Cu$_2$MnSn and
Cu$_2$MnIn, respectively. As shown in Fig.~\ref{f5} good agreement
between the {\it ab-initio} calculations and the experimental data
is also found for these compounds. Similarly to the Cu$_2$MnAl case,
the use of the complex DH potential shows a better performance that
the complex HL one into reproducing the energy difference,
$\Delta$E, among the spectral features. Experimentally, the energy
difference between the main peak (B) and the first negative deep (E)
is 26 eV for both Cu$_2$MnIn and Cu$_2$MnSn compounds. HL
computations yield $\Delta$E = 23.3 and 22.5 eV for the In and Sn
compounds, respectively. These results are improved by using the
complex DH ECP yielding $\delta$E = 25.9 eV and 24.8 eV for
Cu$_2$MnIn and Cu$_2$MnSn, respectively. As observed for Cu$_2$MnAl,
the intensity ratio of the different spectral features in the first
25 eV's of the spectrum is better reproduced by using HL than DH
ECP. Indeed, the weak structure (C) at the high energy side of the
main peak (B) is correctly reproduced by the HL computation. By
contrast, the absorption maxima calculated by using the HL ECP fall
short the experimental ones, specially regarding the energy region
between 30 to 70 eV above the edge, (D to G structures,
Fig.~\ref{f5}).

We have extended our study to the Cu K-edge in the same materials by
following identical procedure as for the Mn K-edge. Also in this
case, we have verified the need of using {\it l}$_{max}$ = 4 as well
as big clusters to reproduce the Cu K-edge experimental spectra.
Best agreement between the experimental data and the
calculations have been obtained by using 169 atoms to generate the
potential and taken into account the scattering contributions of the
first 137 atoms. Therefore, contributions from atoms located within
the first 6.7 {\AA}, 6.9 {\AA} and 6.95 {\AA} around Cu are taken
into account for Cu$_2$MnAl, Cu$_2$MnSn and Cu$_2$MnIn,
respectively.

\begin{center}
\begin{figure}
\includegraphics[scale=0.75]{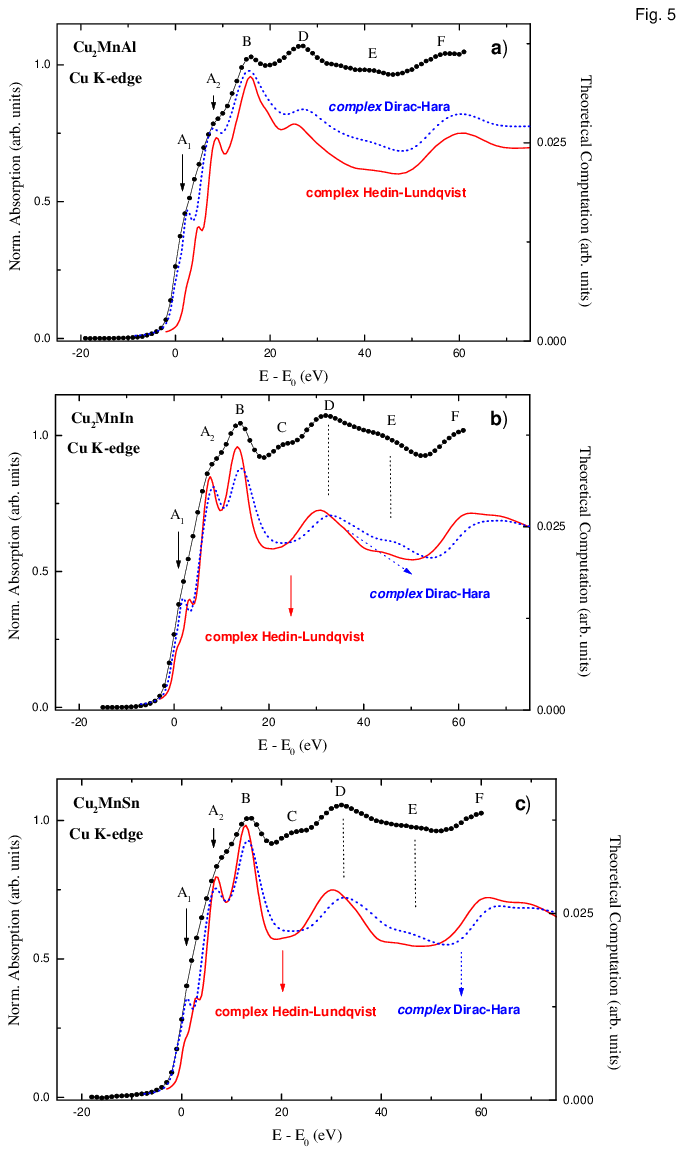}
\caption{(color online) Comparison between the experimental($\circ$)
Cu K-edge of (a) Cu$_2$MnAl, (b) Cu$_2$MnIn and (c) Cu$_2$MnSn, and
the result of the MS computations performed for both complex HL
(red, solid line) and {\it complex} DH (blue, dotted line) ECP
potentials. \label{f6}}
\end{figure}
\end{center}

The experimental Cu K-edge XANES spectra are similar in the three
cases. There are two contributions at the threshold (A$_1$ and
A$_2$) followed by a prominent peak (B) at $\sim$ 15 eV in the three
compounds. In the case of Cu$_2$MnAl, there is a single broad
resonance (D) at about 27 eV above the edge. By contrast, the
spectral profile in this region is different for both Cu$_2$MnSn and
Cu$_2$MnIn compounds. The Cu K-edge XANES of these compounds shows a
small structure (C) on the low energy side of the broad one (D). The
emerging feature is located at $\sim$ 23 eV while the broad D one is
shifted to higher energies ($\sim$ 32 eV) with respect to the
Cu$_2$MnAl case. The theoretical calculations are able of
reproducing this trend. As shown in Fig.~\ref{f6} the computed
spectra for both Cu$_2$MnSn and Cu$_2$MnIn clearly differ from that
obtained for Cu$_2$MnAl in the energy region corresponding to C and
D features, i.e. from 20 to 35 eV above the edge. The best
reproduction of the Cu K-edge experimental spectra is obtained by
using the complex DH potential. Contrary to the Mn K-edge, the
improvement of the complex DH ECP over the HL one concerns both the
intensity ratio and the relative energy separation between the
different spectral features. Indeed, by using the complex HL ECP the
relative intensity of the main contributions to the absorption main
peak is not satisfactory and the contraction of the energy scale
still persists.

 The fact that the agreement between
the theoretical calculations and the experiment is better when using
the complex DH instead of HL potential is noticeable. Indeed, it is
commonly assumed that the HL potential, with its energy dependent
exchange and its imaginary part, is able to reproduce rather
accurately the experimental spectra in the case of metals. By
contrast, several works have reported the improved performance of
the DH potential into accounting for the experimental spectra of
ionic systems, insulators and transition metal oxides
\cite{Gunnella,Itrio,Ti,Cabaret}. This result suggests that the
treatment of the Coulomb hole correlation by using a single
plasmon-pole within the GW approximation \cite{Hedin2} is rather
crude in the high energy region. HL potential is separated into
three parts,
\begin{equation}
  V^{HL}=V^{DH}_{ex}+V_{sex}+V_{ch}
\end{equation}
where $V^{DH}_{ex}$ is the bare exchange potential, Dirac-Hara,
$V_{sex}$ is the screening of exchange potential and $V_{ch}$ is the
Coulomb correlation hole \cite{Hedin,Hatada}. $V^{DH}_{ex}$ is
negative and at high energy asymptotically behaves like, $k^{-2}$.
According to several authors this energy dependence is crucial in
the high-energy part of the absorption spectrum, EXAFS
\cite{Chou,Chou2}. $V_{sex}$ is positive and roughly constant and
then rapidly decreases like $k^{-4}$, while $V_{ch}$, which
describes the repulsion of electrons of the same spin, is negative
and its behavior is $k^{-1}$ at high energy \cite{Hatada}. As
discussed in Ref. \cite{Hedin2} Coulomb correlation hole will lower
the energy of the states and, as a consequence, the position of the
peaks in continuum state is lowered. The Coulomb correlation hole
effect dominates in the higher energy region, in agreement to our
results. Indeed, the comparison of the computations performed by
using DH and HL ECP shows that in the latter case all the peaks in
the high energy region are shifted to a lower energy position. This
result reflects the inadequacy of using a single plasmon-pole within
the GW approximation \cite{Hedin2} to account for the Coulomb hole
correlation in the high energy region.

\section{Summary and Conclusions}

We have presented the detailed {\it ab-initio} computation of the Mn
and Cu K-edge XANES spectra in the case of Cu$_2$MnM (M= Al, Sn and
In) compounds performed within the multiple-scattering framework.

The comparison between the experimental data at the two edges and
the theoretical calculations has demonstrated the need of using
large clusters, $\sim$ 7{\AA} around central photoabsorbing atom,
either Mn or Cu, to reproduce the experimental spectra of these
Heusler alloys.

During the multiple-scattering {\it ab initio} calculations we have
tested different exchange and correlation possibilities to construct
the final state potentials. Contrary to the current expectation for
metallic systems of obtaining the best reproduction of the
experimental XANES spectra by using the Hedin-Lundqvist potential,
the best agreement is obtained, specially in the Cu K-edge case, by
using the Dirac-Hara ECP.

We concluded that Hedin-Lundqvist potential overestimates the effect
of the Coulomb correlation hole in the high energy region resulting
in an unphysical contraction of the computed spectra by lowering in
energy the calculated peak positions. This problem is solved by
adding to the real energy-dependent Dirac-Hara potential the
imaginary part of the Hedin-Lundqvist ECP. In this way, the
inelastic losses of the photoelectron are accounted and both the
relative energy position and the intensity ratio among the different
spectral features are properly accounted by the computations.

\section*{Acknowledgments}

This work was partially supported by the Spanish CICYT (Grant
MAT2005-06806-C04-04). K. Hatada acknowledges a grant from Arag\'on
DGA (research mobility program) and J. Ch. acknowledges a fellow
from the Japanese Society for the Promotion of Science: Invitation
Fellowship Program for Research in Japan. We acknowledge Prof. H.
Maruyama for providing us the experimental spectra.

%
% ************************************* References ***************************
%
%

%

%
%
\end{document}